# Layer-resolved band bending at the *n*-SrTiO$_3$(001)/*p*-Ge(001) interface

Y. Du[1], P.V. Sushko[1], S.R. Spurgeon[2], M.E. Bowden[3], J.M. Ablett[4], T.-L. Lee[5], N.F. Quackenbush[6], J.C. Woicik[6], S.A. Chambers[*,1]

[1]*Physical and Computational Sciences Directorate, Pacific Northwest National Laboratory, Richland, WA 99352, USA*

[2]*Energy and Environment Directorate, Pacific Northwest National Laboratory, Richland, WA 99352, USA*

[3]*Environmental Molecular Sciences Laboratory, Pacific Northwest National Laboratory, Richland, WA 99352, USA*

[4]*Synchrotron SOLEIL, L'Orme des Merisiers, Saint-Aubin, BP 48, 91192 Gif-sur-Yvette, France*

[5]*Diamond Light Source, Ltd., Harwell Science and Innovation Campus, Didcot, Oxfordshire OX11 0DE, England, United Kingdom*

[6]*Materials Measurement Science Division, Material Measurement Laboratory, National Institute of Standards and Technology, Gaithersburg, MD 20899, USA*

The electronic properties of epitaxial heterojunctions consisting of the prototypical perovskite oxide semiconductor, *n*-SrTiO$_3$ and the high-mobility Group IV semiconductor *p*-Ge have been investigated. Hard x-ray photoelectron spectroscopy with a new method of analysis has been used to determine band alignment while at the same time quantifying a large built-in potential found to be present within the Ge. Accordingly, the built-in potential within the Ge has been mapped in a layer-resolved fashion. Electron transfer from donors in the *n*-SrTiO$_3$ to the *p*-Ge creates a space-charge region in the Ge resulting in downward band bending which spans most of the Ge gap. This strong downward band bending facilitates visible-light, photo-generated electron transfer from Ge to STO, favorable to drive the hydrogen evolution reaction associated with water splitting. Ti 2p and Sr 3d core-level line shapes reveal that the STO bands are flat despite the space-charge layer therein. Inclusion of the effect of Ge band bending on band alignment is significant, amounting to a ~0.4 eV reduction in valence band offset compared to the value resulting from using spectra averaged over all layers. Density functional theory allows candidate interface structural models deduced from scanning transmission electron microscopy images to be simulated and structurally optimized. These structures are used to generate multi-slice simulations that reproduce the experimental images quite well. The calculated band offsets for these structures are in good agreement with experiment.

## I. INTRODUCTION

The properties of structurally and compositionally well-defined epitaxial oxide/semiconductor heterojunctions are of considerable fundamental scientific interest, as well as being important in technical areas. In particular, the electronic properties have been of long-standing importance in gate dielectric technology and, more recently, are of potential interest in clean energy applications such as photoelectrochemical water splitting [1]. The prototypical crystalline oxide for fundamental studies of oxide/semiconductor heteroepitaxy is SrTiO$_3$ (STO) [2-15]. STO is a wide gap semiconductor ($E_g$ = 3.25 eV) that is readily doped $n$-type by La$_{Sr}$, Nb$_{Ti}$ and O vacancies ($V_O$). Fabricating crystalline oxide/semiconductor interfaces without amorphous native oxide formation requires a high degree of control over the epitaxial film growth process. It is of ongoing interest to understand the relationship between interface structure and electronic properties, particularly band bending, band alignment and interface state density. In previous investigations of STO/semiconductor heterojunctions, band alignment has typically been measured without looking for and incorporating the effects of band bending. Rather, flat bands have been assumed on both sides of the interface. Yet, band banding, if present, can have a substantial effect on band alignment. The two must be simultaneously detected and quantified in order to accurately elucidate the electronic structure of the interface.

While a significant literature exists for the deposition and properties of simple and complex oxides on Si, relatively less attention has been paid to oxides on Ge. Yet, Ge is of interest for several reasons. Due to the sizeable lattice mismatch between STO and Ge, STO epitaxial films on Ge may exhibit ferroelectric distortions, as have also been reported for STO/Si(001) [9,10]. Indeed, relaxor ferroelectric behavior has been observed for SrZr$_{0.7}$Ti$_{0.3}$O$_3$/Ge(001) [16]. In addition to having substantially higher room-temperature mobilities for both electrons and holes than Si, as well as a smaller bandgap (0.66 eV) to facilitate photovoltaic and photoelectrochemical energy conversion, Ge is more oxidation resistant than Si. The standard free energies of formation of GeO$_2$ and SiO$_2$ are approximately -550 and -910 kJ/mol, respectively, making it easier to deposit oxides on Ge without unwanted native oxide. Nevertheless, deliberate steps must be taken in nucleating the initial interface layers of STO to avoid GeO$_x$ formation. It has been shown that by first incorporating a SrGe$_2$ template layer, structurally coherent interfaces of STO (also SrTi$_{1-x}$Zr$_x$O$_3$) and $p$-Ge(001) can be made without GeO$_x$ formation [13,17], as was first demonstrated for MBE-grown STO on Si(001) using a monolayer of SrSi$_2$ [2]. Provided an atomically abrupt, structurally coherent interface can be formed, the STO/Ge system is a useful testbed for gaining insight into the relationships between structure, composition and functional properties at a prototypical complex oxide/Group IV semiconductor interface.

Here we present a detailed investigation of the junction of *n*-SrTiO$_3$(001) and *p*-Ge(001) prepared by molecular beam epitaxy (MBE). Deposition of epitaxial STO on Ge without GeO$_x$ formation requires the use of low oxygen pressure, resulting in the formation of O vacancies ($V_O$) and *n*-type doping. Epitaxial growth of STO on *p*-Ge thus naturally results in *pn* heterojunction formation. Our interest in this system stems from the fact that Ge is an attractive material for visible light harvesting and water splitting. *Ab initio* calculations reveal that the conduction band minimum (CBM) of Ge is at a higher electron energy than the half-cell potential for the hydrogen evolution reaction (HER) [18]. As a result, photo-excited electrons in Ge could in principle drive the HER. At the same time, the half-cell potential for the oxygen evolution reaction (OER) is predicted to be higher in hole energy than the valence band maximum (VBM), precluding the use of photo-generated holes in Ge to drive the OER. Additionally, these calculations show that the photo-oxidation potential of Ge is nearly degenerate with the VBM. Therefore, Ge may photo-oxidize and the resulting surface would likely trap the photo-generated electrons and prevent the HER from occurring. However, if an epitaxial oxide with a larger bandgap and suitable band alignment can be deposited on Ge to make a trap-free interface, the Ge surface would be chemically protected, thereby allowing electrons generated in Ge to traverse through the oxide to the aqueous solution and drive the HER. We believe that STO is such as oxide. Indeed, we have recently demonstrated that visible light illumination of epitaxial *n*-STO/*p*-Ge(001) heterostructures results in water reduction to evolve H$_2$ gas [1].

Band alignment and the presence of built-in potentials are critically important in realizing the photochemical scenario described above. X-ray photoelectron spectroscopy (XPS) provides the information required to determine the energy landscape in the vicinity of surfaces and heterojunctions. The use of core level and valence band binding energies readily allow band offsets to be measured in a straightforward way, as has been done for years. Built-in potentials can in principle be detected by analyzing core-level broadening. However, surface and interface chemistry can also broaden core-level peaks, thus complicating the determination of built-in potentials. In a previous publication, we noted that both the Ge 3d and Sr 3d line shapes are broader for 5 unit cells (u.c.) STO/*p*-Ge(001) heterojunctions than they are for the pure Ge and STO reference materials [19]. We concluded that the Sr 3d broadening was caused by the presence of a surface layer of Sr(OH)$_2$ whereas the origin of the Ge 3d broadening (interface chemical shift vs. band bending) could not be determined with the data at hand. Here we present new data and analysis that allow us to discriminate between interface chemistry and band bending and build a case for the latter as being the cause of the Ge 3d broadening. The impact of band bending within the Ge on accurate determination of band offsets is explored and found to be substantial. Additionally, by employing scanning transmission electron microscopy (STEM), first principle modeling and image simulations, we determine that two structural motifs coexist at the interface. We find that the calculated band offsets for these two structures are the same, and match experiment.

**II. METHODS**

SrTiO$_3$ epitaxial films with thicknesses of 3, 6, 9 and 32 u.c. were deposited using MBE on clean, Ga-doped *p*-Ge(001)-(2x1) substrates for which $\rho$ = ~2 $\Omega$-cm and $[h^+]$ = ~1.2×10$^{-15}$ cm$^{-3}$, as measured by the Hall effect. All films were prepared by first depositing 0.5 monolayers (ML) Sr at 400°C to create an ordered (3×1) oxidation-resistant monolayer, followed by co-deposition of 2.5 ML Sr and 3.0 ML Ti with the substrate at ~25°C in 4×10$^{-7}$ Torr O$_2$. After pumping out the O$_2$, the amorphous 3 u.c. film was recrystallized by annealing in high vacuum at temperatures up to ~630°C for a few minutes. These steps resulted in a crystalline 3 u.c. STO template layer with no GeO$_x$ at the interface. Following this procedure for the first 3 u.c., the remainders of the 6, 9 and 32 u.c. films were co-deposited using progressively higher substrate temperatures as the film thickness increased, in order to maximize the extent of crystallization without promoting Ge oxidation. For the 6 u.c. and 9 u.c. films, unit cells 3 – 6 were deposited at a substrate temperature of 250°C and an O$_2$ pressure of 2×10$^{-6}$ Torr. For the 9 u.c. film, the first 6 u.c. were synthesized as described above, and unit cells 6 – 9 were co-deposited at 400°C. Following growth, the 6 and 9 u.c. films were annealed in vacuum at 550°C for improved crystallization. For the 32 u.c. film, the first 3 u.c. were deposited as described above, and unit cells 4 – 32 were grown by co-deposition while ramping the substrate temperature from 550°C to 600°C and the O$_2$ pressure from 4×10$^{-7}$ Torr to 2×10$^{-6}$ Torr. No post-annealing in high vacuum was carried out on this sample. For all films, the growth rate was ~40 sec per u.c.

Core-level and valence band spectra yielding information on chemical speciation, band bending and valence band offsets (VBO) were measured *in situ* using high-energy-resolution XPS for the 3 – 9 u.c. samples. *Ex situ* hard x-ray photoemission spectroscopy (HAXPES) with ~6 keV x-rays yielded complimentary information for the 32 u.c. sample. *In situ* XPS measurements were performed at normal emission using a Scienta Omicron R3000 analyzer [20] and a monochromatic AlK$\alpha$ x-ray source ($h\nu$ = 1487 eV) with an energy resolution of ~0.5 eV. The binding energy scale was calibrated using the Ag 3d$_{5/2}$ core level (368.21 eV) and the Fermi level from a polycrystalline Ag foil. HAXPES measurements were made at the Diamond Light Source (UK) on the I09 Surface and Interface Structural Analysis beamline at an x-ray energy of 5930 eV. This beamline includes a Scienta Omicron EW4000 high-energy analyzer [20] with an overall energy resolution of < 250 meV at $h\nu$ = ~6 keV. The binding energy scale was calibrated using the Au 4f and Au 4p core levels, along with the Fermi edge of a gold foil. The x-ray angle of incidence was 5º off the surface plane, and the photoelectron detection angle was 5º off normal.

Cross-sectional STEM samples were prepared using a FEI Helios NanoLab DualBeam Focused Ion Beam (FIB) microscope [20] and a standard lift out procedure along the Ge [110] zone axis, with initial cuts made at 30 kV / 2° and final polishing at 1 kV / 3° incidence angles. High-angle annular dark field (STEM-HAADF) images were collected on a JEOL ARM-200CF microscope [20] operating at 200 kV, with a convergence semi-angle of 20.6 mrad and a collection angle of 90-370 mrad. To minimize scan artifacts and improve signal-to-noise, drift-corrected images were prepared using the SmartAlign plugin; [21] for this, a series of 10 frames at 1024 x 1024 pixels with a 2 μs px$^{-1}$ dwell time and 90º rotation between frames was used. The frames were up-sampled 3x prior to non-rigid alignment. The images were subsequently processed using a lattice enhanced filter [22]. Full multi-slice image simulations were conducted with the PRISM code [23] for the structures from our first-principles calculations. Simulations were performed using a 1 × 3 tiling for crystal thicknesses of 1, 13, 38, and 75 u.c., corresponding to ~ 0.8, 10, 30, and 60 nm, respectively, as shown in the supplemental information. Imaging parameters were matched to the experiment and a 0.05 Å pixel sampling, 2 Å slice thickness, and 10 frozen phonon passes were used for the final simulations. From these simulations, the sample thickness is estimated to be 10–30 nm.

Modeling of the STO/Ge heterojunction was carried out using a periodic slab of the form STO/Ge/STO, where the two interfaces are mirror images. The lateral cell includes four Ge atoms per atomic plane and, accordingly, four unit cells of STO. The Ge portion of the slab includes 19 atomic planes and each STO film contains 7 or 8 atomic planes depending on the details of the structure. The in-plane lattice parameters were fixed at the value pre-calculated for bulk Ge (8.069 Å); the out-of-pane parameter was selected to provide ~20 Å vacuum gap between the slabs. The calculations were carried out using the Vienna *Ab initio* Simulation Package (VASP) [24,25]. The projector-augmented wave was used to approximate the electron-ion potential [26]. In all cases we used the PBEsol density functional [27] to optimize the geometrical structures. The PBEsol + *U* approach with a rotationally invariant Hubbard *U* correction ($U$ = 8.5 eV) was applied to Ti 3d states [28] and used for post-optimization calculation of the density of states (DOS). The plane wave basis set cutoff was set to 500 eV. A Γ-centered 2×2×1 Monkhorst-Pack grid were used for the structure optimization and 8×8×1 grid was used for calculations of the DOS.

## III. RESULTS AND DISCUSSION

### A. Film and Interface Structure

The in-plane lattice mismatch between STO(001) and Ge(001), $\{[a_{STO} - a_{Ge}\cos(45)] / a_{Ge}\cos(45)\}$, is -2.3% after accounting for the 45º rotation of the film relative to the substrate about [001] required by epitaxial registry. Reflection high-energy electron diffraction (RHEED) patterns for the clean Ge substrate and the four epitaxial STO films (Fig. 1) show intense, unmodulated streaks with low background, revealing heteroepitaxial nucleation with the expected 45º rotation about the surface normal, as well as smooth film surfaces. In-plane lattice parameters were extracted from the RHEED streak spacings and these, along with an x-ray diffraction (XRD) direct space map (DSM) for the 32 u.c. film, are shown in Fig. 2. In-plane lattice parameters were determined by analyzing RHEED line profiles measured in the [100] and [110] zone axes near the tops of the patterns. Clean Ge(001) and STO(001) single crystals were used as standards. Individual rods were fit to Voigt functions after linear background subtraction to determine streak spacings. The standards were used to convert the differences between streak positions for symmetry-equivalent rods on opposite sides of the specular beams to in-plane lattice parameters. These measurements were in turn averaged over the two azimuths to yield in-plane lattice parameter vs film thickness.

To quantify the structure of the 32 u.c. STO/Ge interface, we employ a novel analysis method in which we interpret high-resolution STEM-HAADF images using multi-slice simulations iteratively refined based on *ab initio* calculations. This approach has been successfully demonstrated in other systems, such as two-dimensional materials [29] and layered oxides [30], but it has been used far less to interpret complex oxide/semiconductor interfaces, such as $SrTiO_3$/Si [31].

Here we first identified a range of possible interface structures, ranked by geometric compatibility and lowest energy cost based on *ab initio* calculations. The most promising candidates were selected as inputs for multi-slice simulations; these, in turn, were compared to direct images of the interface. The process was then iterated, taking into account discrepancies between the theoretical models and the measured images, ultimately yielding excellent agreement for both structure and calculated electronic properties. Representative cross-sectional images shown in Fig. 3 confirm that the film is epitaxial, with a sharp film-substrate interface as determined through chemical mapping described elsewhere [1]. We note the presence of some defects within the film, principally antiphase boundaries (APB) (shown in the center of the figure). We also observe two predominant interface configurations that we have termed "1×1 Ge" and "2×1 Ge," shown in detail in the lower half of Fig. 3. A third configuration appears in the immediate vicinity of the APB, but it is likely the result of lattice distortion and strain associated with the APB, so we do not discuss it further. The 1×1 Ge and 2×1 Ge HAADF structures were used to create trial models that were subsequently optimized in DFT, input into multislice simulations, and used to down select to two atomistic models for the experimentally observed 1×1 Ge and 2×1 Ge structures. These structures and their electronic properties are discussed below in conjunction with Figs. 9 & 10.

For the 1×1 Ge structure, we find that interface Ge atoms bind to the Ti atoms. The formation of these Ge–Ti bonds is facilitated by four oxygen vacancies located in the interface $TiO_x$ plane, each donating two electrons to saturate two dangling bonds of each interfacial Ge. In addition, two oxygens displace from the interfacial $TiO_x$ to the neighboring $Sr_2O_2$ plane, thus creating a $Ge_4 - Ti_4O_2 - Sr_2O_4 - Ti_4O_8$… layer sequence. Assuming formal ionic charges, including a –1 charge for the saturated Ge dangling bonds, the resulting charge distribution in the near-interface $Ge/TiO_x/SrO_y$ planes is –8/+12/–4 per lateral cell. Since the distance between the Ge and $TiO_x$ planes is approximately half of the distance between the $TiO_x$ and $SrO_y$ planes, this configuration corresponds to well-compensated interface dipole. Because the STEM-HAADF signal from an atom with atomic number $Z$ is proportional to $Z^{\sim 1.7}$, this decrease in occupancy accounts for the reduction in scattered intensity at the interface and produces an excellent match with the experimental image, as seen in Fig. 3.

For the 2×1 Ge structure, we confirm the presence of Ge dimers at the interface into a (2×1) reconstruction, as has been reported elsewhere [32]. The reconstructed substrate surface is matched to a half monolayer of Sr, yielding the sequence $Ge_4 - Sr_2 - Ti_4O_8 - Sr_4O_4$. The multi-slice simulation again shows a striking correspondence to the experimental image. We note that the experimental image shows occasional signal intensity where the dimers converge within the $Sr_2$ layer (marked by arrow). However, because these features appear to be random, we have not attempted to model them. Taken together, our STEM results reveal an abrupt film-substrate interface consisting of two dominant structural motifs that are well-described by first-principles calculations and can be used to calculate the band alignment for the purpose of comparison with experiment, as described below.

### B. Electronic properties

In the absence of band bending, band alignment can be accurately determined in a straightforward way by matching the measured heterojunction valence band (VB) spectrum to a linear combination of appropriately weighted spectra for the pure component materials separated in energy by the valence band offset (VBO). We show in Fig. 4 such an analysis for STO/Ge. VB spectra for the four heterojunctions, along with the best fits to linear combinations of spectra for pure, bulk *p*-Ge(001) and 1 at. % Nb-doped STO(001). The weighting factor and valence band offset were free parameters in each case. The best fits occur for a VBO ($\Delta E_V$) of 3.2(1) eV for 3, 6 and 9 u.c., and 3.1(1) eV for 32 u.c. It is important to note, however, that this method averages over all layers within the probe depth and thus tacitly assumes that the bands are flat. In what follows, we use core-level line shapes to simultaneously probe band bending and band alignment. In this analysis, it is essential to know how deep we are probing. We thus seek accurate values for the electron attenuation lengths, or EAL ($\lambda$).

The VB spectral data can be used to determine $\lambda$ appropriate for the STO/Ge system at kinetic energies of ~1.5 keV. The weighting factors mentioned in the previous paragraph for the pure Ge and STO VB spectra with STO film thickness $t_{STO}$ should be of the order of $\exp\left(-\frac{t_{STO}}{\lambda}\right)$ and $1 - \exp\left(-\frac{t_{STO}}{\lambda}\right)$, respectively, provided the heterojunction and pure component spectra are measured with the same spectrometer settings and x-ray fluxes. As a result, the best-fit weighting factors for 3, 6 and 9 u.c. can be used to determine EALs for AlK$\alpha$ excitation by employing the formula,

$$A_{STO/Ge} \cong A_{Ge} \exp\left(-\frac{t_{STO}}{\lambda}\right) + A_{STO}\left[1 - \exp\left(-\frac{t_{STO}}{\lambda}\right)\right] \qquad (1)$$

where $A$ represents integrated VB spectral areas. Solving eqn. (1) leads to $\lambda$ values of 2.2, 2.1 and 2.0 nm for 3, 6 and 9 u.c., respectively. With these values, the weighting factors from the solutions to eqn. 1 match those resulting from the best fits shown in Fig.4 to within ~10%. We thus conclude that $\lambda = 2.1 \pm 0.2$ nm is appropriate for VB measurements and shallow core levels excited at $h\nu = 1.5$ keV for epitaxial STO on Ge. We estimate $\lambda$ for $h\nu = 6$ keV by using the universal $E_k^{\sim 0.8}$ dependence of $\lambda$ on the photoelectron kinetic energy ($E_k$) well above the minimum at ~100 eV [33,34]. In this case, $\lambda_{6\,keV} \cong 2.1(6/1.5)^{0.8} = 6.0$ nm.

We also consider calculated EALs based on the model of Jablonski [35,36] in which inelastic mean free paths determined by solving a modified Bethe equation are corrected for elastic scattering and x-ray polarization. Doing so yields $\lambda$ values of ~2.4 nm for both Ge and STO at 1480 eV, and 6.0 nm for STO at 6 keV, in good agreement with our experimental values. This method also generates a value of 8.6 nm for Ge at 6 keV, which we use in our Ge 3d line shape analysis for 32 u.c. STO on Ge.

A representative set of core-level (CL) spectra are shown in Fig. 5. Here we overlay the heterojunction spectra with reference spectra for $p$-Ge(001) and various STO(001) specimens. A bulk 1 at. % Nb-doped STO(001) was used for the O1s, Ti 2p and Sr 3d HAXPES. Homoepitaxial STO(001) with a TiO$_2$ termination was used for O 1s and Sr 3d XPS, and SrO-terminated homoepitxial STO was used for Ti 2p XPS. The reason for using these homoepitaxial STO film surfaces for Ti 2p and Sr 3d was to minimize line shape asymmetries or separate peaks resulting from surface chemical shifts and/or surface reactivity by insuring that the atom of interest (Ti or Sr) is not present in the surface atomic layer [19]. We use the TiO$_2$-terminated homoepitaxial film surface for the O 1s line shape because the spectrum for the analogous SrO-terminated surface often contains a surface hydroxide feature.

The layout in Fig. 5 is designed to illustrate changes in line shapes accompanying heterojunction formation relative to those of the pure materials. If the substrate had oxidized during STO film growth, GeO$_x$ would appear as a broad feature centered ~3 eV to higher binding energy from the centroid of the lattice spin-orbit (SO) Ge 3d doublet (see Fig. 6). The absence of such a feature indicates that GeO$_x$ formation did not occur during interface formation. However, in all cases, the Ge 3d spectra exhibit an asymmetric broadening to higher binding energy relative to pure *p*-Ge(001), and the extent of broadening increases with increasing STO film thickness.

The Ge 3d asymmetries can be rationalized by fitting the spectra using two SO doublets, one assigned to pure Ge away from the interface and the other (a weaker doublet) to Ge atoms near the interface and shifted 0.3 eV to higher binding energy due to their interaction with STO. However, this interface model can be ruled out based on intensity considerations. When the Ge 3d spectra are fit this way (not shown), the intensity ratios of the weaker, higher-binding-energy SO pair to the more intense, lower-binding-energy SO pair differ significantly from what is expected for an interfacial Ge layer at any STO film thickness. The ratio of Ge 3d intensity associated with an interface phase to that for bulk Ge in the lattice below the interface can be approximated by

$$\frac{A_{\text{Ge int}}}{A_{\text{Ge latt}}} = \frac{\sum_{j=1}^{n} \exp\left[\frac{-(t_{\text{STO}} + jd_{\text{Ge}})}{\lambda}\right]}{\sum_{k=1}^{\infty} \exp\left\{\frac{-[t_{\text{STO}} + (n+k)d_{\text{Ge}}]}{\lambda}\right\}} = \frac{\sum_{j=1}^{n} \exp\left(\frac{jd_{\text{Ge}}}{\lambda}\right)}{\sum_{k=1}^{\infty} \exp\left\{\frac{-[(n+k)d_{\text{Ge}}]}{\lambda}\right\}} \qquad (2)$$

where *n* is the number of layers in the interface phase. The STEM images in Fig. 3 clearly show that the interface is atomically abrupt which means that *n* = 1. As a result, the ratio given by eqn. 2 with $\lambda$ = 2.1 nm is 0.069, independent of STO film thicknesses. However, the actual intensity ratios measured with AlK$\alpha$ x-rays are 0.025, 0.054 and 0.11 for the 3, 6 and 9 u.c. heterojunctions, respectively. Likewise, for the 32 u.c. heterojunction probed with 6 keV x-rays, eqn. 2 predicts this ratio to be 0.016 for $\lambda$ = 8.6 nm, whereas the experimental HAXPES ratio is 0.38.

We also see in Fig. 5 that the Sr 3d spectra exhibit an asymmetric broadening to higher binding energy relative to the TiO$_2$-terminated, homoepitaxial STO reference spectrum. This asymmetry can be accounted for by using a second SO doublet shifted ~0.9 eV to higher binding energy from the lattice doublet. The weaker SO pair may be due to a surface-bound SrO$_x$ species [37]. Alternatively, this asymmetry could be due to a chemical shift at the buried interface. In any event, band bending in the STO can be ruled out as the cause of the Sr 3d asymmetry because the O 1s and Ti 2p$_{3/2}$ peaks do not show the same asymmetry, and it is well known that band bending broadens and shifts core-level features to the same extent for all atoms in the material under study.

The Ti $2p_{3/2}$ spectra reveal symmetric broadening at 3 u.c. and a very slight asymmetry to lower binding energy relative to the line shape for SrO-terminated homoepitaxial STO(001) for 6 u.c. and 9 u.c. spectra. This asymmetry requires a peak shifted ~1.0 eV to lower binding energy from that for lattice $Ti^{4+}$ in order to get a good fit. This feature is indicative of $Ti^{3+}$ which results from $V_O$ formation associated with deposition in low $O_2$ pressure to avoid Ge oxidation. The HAXPES spectrum for the 32 u.c. film is narrow and symmetric, with a FWHM of 0.80 eV, the same value measured for bulk STO(001), and $Ti^{3+}$ is not detected.

The O 1s spectra consist of single peaks with the same full widths at half maximum (FWHM) value as those measured for homoepitaxial STO(001) [19]. We note that the raw 32 u.c. heterojunction spectrum exhibits a weaker feature to higher binding energy from the lattice peak assigned to surface contamination (organics and water) resulting from the through-air transfer from the MBE system to the HAXPES chamber [38,39]. This feature has been removed by subtraction in Fig. 5. No such feature is seen in spectra measured *in situ* for the thin films (3, 6, and 9 u.c.). This result allows us to conclude that the high-binding-energy features in the Sr 3d spectra are not due to $Sr(OH)_2$, as was seen in an earlier study [19].

The dashed vertical lines in Fig. 5 denote the Ge $3d_{5/2}$, Sr $3d_{5/2}$ and Ti $2p_{3/2}$ binding energies that would indicate the Ge VBM and STO CBM being degenerate with the Fermi level ($E_F$). The positions of these lines are based on the measured energy differences between the core peaks and the VBMs for pure *p*-Ge(001) and *n*-STO(001) ($\Delta E_{CL-VBM}$, see discussion below) and the STO bandgap (3.25 eV). From these, we see that the STO CBM is within 0.1 eV of $E_F$ for all heterojunctions, establishing that the STO is *n*-type. Moreover, the absence of uniform asymmetry in the Sr 3d, Ti 2p and O 1s spectra relative to those for flat-band STO standards reveals that there is negligible band bending in the STO films.

The Ge 3d spectra also indicate that when averaged over all layers within the probe depth, the Ge VBM is quite close to $E_F$ for all heterojunctions. However, the broadening and asymmetry may be signaling the presence of band bending within the Ge. The asymmetry to *higher* binding energy is consistent with *downward* band bending; layers closest to the interface would exhibit higher binding energies than those deeper in, and the deeper layers would converge to a constant value characteristic of the part of the probe depth below the depletion zone and close to $E_F$. In order to test for band bending, we fit the experimental Ge 3d heterojunction spectra, $I_{expt}(\varepsilon)$, to linear combinations (LC) of model functions which are themselves fits to spectra measured for pure *p*-Ge(001). As seen in Fig. 6, the line shapes of these reference spectra reveal that they are symmetric due to the absence of band bending and are thus suitable to represent individual layers within the depletion zones of the heterojunctions. Looking first at the spectra excited with AlKα x-rays for clean *p*- and *n*-Ge(001), we see that the VBM is within a few hundredths of an eV of the Fermi level. This result is consistent with previous angle-resolved photoemission, scanning tunneling spectroscopy, and transport studies which conclude that the Fermi level is pinned near the VBM at (001)-oriented surfaces of both *n*- and *p*-type Ge(001) [40-42]. The Ge 3d spectra are well fit to pairs of Voigt functions and show no asymmetry which would be visible if band bending was occurring. In contrast, the Fermi level is near mid gap in the spectrum excited with 6 keV x-rays, for which the sample was not cleaned prior to measurement (thus the $GeO_2$ feature). Similar to the AlKα excited spectra, however, the Ge 3d spectrum excited at 6 keV is well fit using a pair of Voigt functions and shows no asymmetry, indicating a flat-band state. Indeed, solution of charge neutrality and Poisson's equations for Ga-doped Ge with an acceptor density of $1 \times 10^{15}$ cm$^{-3}$ reveals that the bulk Fermi level is 0.22 eV above the VBM. There is thus ~0.1 eV of band bending, but the depletion width is 420 nm, more than an order of magnitude larger than the HAXPES probe depth.

Model functions that produce accurate fits to the spectra in Fig. 6 are assigned to all layers in the various heterojunctions and their intensities are attenuated according to their depths below the interface (*z*) using an inelastic damping factor of the form exp(-*z*/λ). The fitting algorithm starts by randomly generating binding energies for spectra associated with the different layers within the probe depth. These energies are then sorted from highest to lowest and are assigned to the layers to ensure that the binding energy at maximum intensity, $\varepsilon_{max}(j)$, is a monotonic function of depth, as expected for a space charge region. This peak binding energy set $\{\varepsilon_{max}(j)\}$ is a measure of the potentials within the layers because core-level binding energies, like VBMs, scale with electrostatic potential. The spectra for all layers were then summed to generate a trial simulated heterojunction spectrum, $I_{sim}(\varepsilon)$. Optimization of the binding energies $\varepsilon$ proceeds so as to minimize a cost function, defined as

$$\chi = \sqrt{\frac{1}{n}\sum_{i=1}^{n}[I_{exp}(\varepsilon_i) - I_{sim}(\varepsilon_i)]^2 + p\sum_{j=1}^{m}[\varepsilon_{max}(j) - \varepsilon_{max}(j+1)]^2} \qquad (3)$$

The first term of $\chi$ is a root-mean square deviation that quantifies the goodness of the fit between the measured and simulated spectra; the sum is over the number of discrete energies in the experimental spectrum. The second term is a sum over layers (*j*) designed to minimize discontinuities in the potential ($\varepsilon_{max}$) with depth. The weighting factor *p* is included to scale the influence of the continuity condition relative to that of the spectrum fitting condition. We find that modest values of *p* (~0.05 – 0.07) are adequate to impose continuity of the potential without sacrificing quality in the spectral fit. Following the initial assignment of the binding energies to the various layers, these energies are subjected to incremental random changes and reordering and the process is repeated until $\chi$ is minimized. We have carried out several fits using different signs for the potential gradients in both the STO and Ge, different numbers of layers within the probe depth subject to energy variation ($m$), and different values of the maximum increment per step, in order to adequately sample the large phase space of the potential energy profiles.

The best fits are shown in Fig. 7. On the left side of each panel, the families of spectra are shown for all layers within the probe depths, which we define as the depth at which a photoemission signal from an atom on the surface would be 99.9% attenuated. In carrying out these fits, we found that the Ge 3d binding energy converged to a constant value at ~30 layers for 3, 6 and 9 u.c. STO/Ge probed at $h\nu$ = ~1.5 keV, and at ~120 layers for 32 u.c. STO/Ge probed at $h\nu$ = ~6 keV. Also shown in Fig. 7 as white dashed lines are the Ge $3d_{5/2}$ binding energies that would be measured if $E_F$ was at the VBM and CBM, given by $\Delta E_{Ge3d5/2 - VBM}$ and $\Delta E_{Ge3d5/2 - VBM} + E_g$, respectively, where $E_g$ = 0.66 eV for Ge. In each case, a good fit could be obtained only if we model the bands as bending downward as the interface is approached from the bulk of the Ge; upward band bending did not yield a good fit in any case. The VBM is slightly lower in energy than $E_F$ below the depletion zone for all four heterojunctions. The Ge CBM at the interface is near mid gap for 3, 6 and 9 u.c. but at $E_F$ at the interface for the 32 u.c. heterojunction. On the right side of each panel, we show the sums of spectra from all layers overlaid with the experimental heterojunction spectra. The fits are excellent in all cases, indicating that downward band bending profiles that span some or all of the Ge gap, depending on depth, account very well for the measured spectra. The smaller extents of band bending and smaller depletion widths for 3, 6 and 9 u.c. STO/Ge excited with AlK$\alpha$ x-rays compared to 32 u.c. STO/Ge excited with hard x-rays are consistent with significantly more total electron transfer from STO to Ge when the STO film thickness is larger.

We combine Ge $3d_{5/2}$ binding energies for the interface layers with Ti $2p_{3/2}$ and Sr $3d_{5/2}$ binding energies averaged over all layers to determine VBOs ($\Delta E_V$) based on the CL method described elsewhere [19]. The results are shown in Table 1. The $\Delta E_{CL-VBM}$ values for bulk Nb:STO(001) and $p$-Ge(001) required to determine $\Delta E_V$ with AlK$\alpha$ x-rays averaged over several crystals are 130.49(4), 455.87(4) and 29.34(4) eV for Sr $3d_{5/2}$, Ti $2p_{3/2}$ and Ge $3d_{5/2}$, respectively. The analogous HAXPES values averaged over two crystals are 130.32(4), 455.74(4) and 29.35(4) eV. As seen in Table 1, the VBOs and CBOs determined using: (a) Sr $3d_{5/2}$ & Ge $3d_{5/2}$, and, (b) Ti $2p_{3/2}$ & Ge $3d_{5/2}$ are within experimental error for each heterojunction, and $\Delta E_V$ values averaged over the two pairs of CLs are also within experimental error for 3, 6 and 9 u.c. The VBO and CBO are somewhat smaller at 32 u.c. presumably as a result of more interfacial charge transfer at this larger thickness. Moreover, the VBOs resulting from modeling the band bending in order to determine the Ge $3d_{5/2}$ binding energies directly at the interface differ from those calculated by averaging over all layers using either core levels or valence bands (Fig. 4) by several tenths of an eV. This result indicates the critical importance of quantitatively mapping the band bending in the quest for accurate band offsets.

We show in Fig. 8 an energy diagram for the 32 u.c. heterojunction based the spectral data discussed above. The Ge VBM profiles (red solid circles) were obtained directly from the Ge $3d_{5/2}$ binding energies resulting from the fit shown in Fig. 7d by subtracting $\Delta E_{Ge3d5/2\text{-}VBM}$. The Ge CBM profiles (green solid circles) are the VBM profiles less the Ge band gap (0.66 eV). Likewise, the STO VBM (red line) was obtained from the Sr $3d_{5/2}$ and Ti $2p_{3/2}$ binding energies in Table 1 by subtracting the appropriate $\Delta E_{CL\text{-}VBM}$ values and averaging, and the STO CBM (green line) is the VBM less the STO band gap (3.25 eV). The Ge CBM is at the Fermi level at the interface. The bands bend upward going away from the interface and the Ge VBM converges to a value quite close to the Fermi level at a depth of ~15 nm below the interface. The absence of measurable band bending in the STO is not surprising in light of the large difference in dielectric constant between Ge ($k$ = 16) and epitaxial STO films ($k$ = ~68) [2-15]. Poisson's equation requires continuity of the electric displacement across the interface, which is equivalent to

$$\left(\frac{\partial V}{\partial z}\right)_{STO} = \frac{\varepsilon_{Ge}}{\varepsilon_{STO}}\left(\frac{\partial V}{\partial z}\right)_{Ge} \qquad (4)$$

As seen in Fig, 8, the potential gradient on the Ge side of the interface is ~0.025 V/Å which means that the gradient on the STO side is ~0.006 V/Å, resulting in a built-in potential of only ~0.05 eV across the first two u.c. of the STO. This value is below the band bending detection limit afforded by core-level peak width analysis.

The sign of the band edge gradient in the Ge space-charge region seen in Figs. 7 & 8 indicates that electrons flow from STO to Ge as a result of interface formation. Significantly, we note that neither the potential gradients nor the band offsets can be known *a priori* for perovskite/Group IV semiconductor interfaces due to the complexities of interface structure and composition. Kolpak and Ismail-Beigi [43] have carried out first-principles calculations for a range of structures that may be found at the STO/Si(001) interface and have shown that although the details of the band alignment and band bending vary, a fixed dipole universally forms due to charge transfer from Si to O in the STO directly at the interface. The same is in principle true for STO/Ge. However, the presence of both $V_O$ donors in actual epitaxial films grown on Group IV semiconductors, and Ga acceptors in the present Ge substrates, result in additional degrees of electronic freedom. At the simplest level, the direction of the built-in electric field would be indicated by the values of the work functions (Φ) of *p*-Ge(001) and *n*-STO(001), and the band alignment would be dictated by the difference in the electron affinities. However, the work function and the electron affinity of *n*-STO(001) depend on surface termination. Our ultraviolet photoemission (UPS) measurements for SrO- and $TiO_2$-terminated *n*-STO(001), as well as for *p*-Ge(001)-2×1, yield Φ values of 3.4 eV, 4.7 eV and 4.7 eV, respectively, after correction of band bending [44]. The corresponding electron affinities are 3.4 eV, 4.7 eV and 4.3 eV. In light of the coexistence of the two interface structural motifs seen in Fig. 3, neither of which is purely *p*-Ge/$TiO_2$-SrO-$TiO_2$-SrO… or *p*-Ge/ SrO-$TiO_2$-SrO-$TiO_2$…, we cannot predict which way the bands will bend on this basis, or what the band offsets will be. The analysis described above thus provides unique and valuable insights into the signs and magnitudes of the built-in potentials and band offsets.

Additional insight can be gained by coupling the detailed STEM images in Fig. 3 with first-principles modeling. *Ab initio* calculations reveal that the band alignment is the same for the 1×1 Ge (Fig. 9a) and 2×1 Ge (Fig. 10a) interface structures (see STEM images and simulations in Fig. 3), and are in good agreement with the experimental results. We show in Figs. 9 b & c and 10 b & c layer-projected densities of states (DOS) for these structures without (panels b) and with (panels c) $V_O$ in the STO near the interface on each side of the simulation slab. The dashed vertical lines mark the positions of the top of the Ge VB in each case. No attempt was made to dope the Ge *p*-type because it is not possible to accurately simulate Ga acceptor concentrations of $10^{15}$ cm$^{-3}$ without unworkably large supercells. As a result, there is no band bending apparent in the theoretical simulations.

In Fig. 9b, the layer-projected DOS for the 1×1 Ge structure shows that the top of the Ge VB falls near the middle of the STO band gap. In the case of oxygen deficient STO for this structure (Fig. 9c), the two $V_O$, one at each interface, are most stable in the 3$^{rd}$ atomic plane from the interface. These $V_O$ pin the Fermi level at the bottom of the STO CB. The electron charge redistribution induced at the interface by $V_O$ changes the STO/Ge band offset such that the top of the Ge VB is degenerate with the bottom of the STO CB, in agreement with experiment, if the band bending is not considered. In the case of the 2×1 Ge structure (Fig. 10), the Ge VB also aligns with the STO CB in nominally stoichiometric STO, and $V_O$ formation in the STO does not change this band alignment, also in agreement with experiment in the absence of band bending.

Finally, we note that the electric field on the Ge side of the interface is expected to facilitate photo-generated electron drift into the STO and ultimately into an electrolytic solution in contact with the STO. The photo-reduction potential of Ge is higher in electron energy than the HER half-cell potential [18]. As a result, electrons generated by light absorption in Ge will preferentially flow downhill and can drive the HER in water photoelectrolysis, as observed under cathodic bias [1].

**IV. SUMMARY**

We have investigated the structural and electronic properties of MBE-grown $n$-STO/$p$-Ge(001) heterojunctions using a combination of experimental and theoretical methods. We have explored the previously unexamined connection between band alignment and band bending at the general class of STO/Group IV semiconductor interfaces. Two distinctly different interface structural motifs exist on opposite sides of antiphase domain boundaries in the STO films. First-principles modeling shows that the calculated valence band offsets associated with these structures are the same, and in agreement with experiment. A newly developed method for analyzing high energy resolution core-level x-ray photoemission line shapes and binding energies has uncovered new insights into the presence of built-in potentials, and the effect of these potentials on the associated band offsets. This new approach should be useful for a wide variety of heterostructures involving complex oxides and other kinds of semiconductors for which experimental information on band alignment and built-in potentials is sought to complement simulated energy diagrams, such as those from Schrodinger-Poisson modeling. Our results for the $n$-STO/$p$-Ge(001) heterojunction indicate that electron transfer from STO to Ge during interface formation results in downward band bending across the gap of $p$-Ge, with the Ge conduction band edge becoming degenerate with the Fermi level at the interface for STO film thicknesses of ~10 nm or greater. This built-in potential is expected to facilitate photo-generated electron drift from Ge to STO, a useful feature in photoelectrochemical applications. We also show that the valence band offset is reduced by ~0.4 eV when this trans-gap band bending is included in the analysis compared to when it is ignored, thus establishing the importance of simultaneously determining band alignment and band bending in order to accurately map out the electronic structure of complex heterointerfaces.

## ACKNOWLEDGEMENTS

This work was supported by the U.S. Department of Energy, Office of Science, Division of Materials Sciences and Engineering under Award #10122. The PNNL work was performed in the Environmental Molecular Sciences Laboratory, a national scientific user facility sponsored by the Department of Energy's Office of Biological and Environmental Research and located at PNNL. We thank Diamond Light Source for access to beamline I-09 (SI17449-1) that contributed to the results presented here. S.A.C. acknowledges helpful conversations with Profs. Joe Ngai, Kelsey Stoerzinger and Bharat Jalan. S.R.S. thanks Dr. Colin Ophus for useful discussions about multi-slice simulations.

*Author to whom correspondence should be addressed. Electronic address: sa.chambers@pnnl.gov

# REFERENCES

[1] K. A. Stoerzinger, Y. Du, S. R. Spurgeon, L. Wang1, D. Kepaptsoglou, Q. M. Ramasse, E. J. Crumlin, S. A. Chambers, MRS Comm. **8**, 446 (2018).

[2] R. A. McKee, F. J. Walker, M. F. Chisholm, Phys. Rev. Lett. **81**, 3014 (1998).

[3] R. A. McKee, F. J. Walker, M. F. Chisholm, Science **293** (2001).

[4] R. A. McKee, F. J. Walker, M. B. Nardelli, W. A. Shelton, G. M. Stocks, Science **300** (2003).

[5] H. Li, X. Hu, Y. Wei, Z. Yu, X. Zhang, R. Droopad, A. A. Demkov, J. Edwards, K. Moore, W. Ooms, J. Kulik, P. Fejes, J. Appl. Phys. **93** (2003).

[6] F. S. Aguirre-Tostado, A. Herrera-Gomez, J. C. Woicik, R. Droopad, Z. Yu, D. G. Schlom, P. Zschack, E. Karapetrova, P. Pianetta, C. S. Hellberg, Phys. Rev. B **70**, 201403 (2004).

[7] J. C. Woicik, H. Li, P. Zschack, E. Karapetrova, P. Ryan, C. R. Ashman, C. S. Hellberg, Phys. Rev. B **73**, 024112 (2006).

[8] P. Ryan, D. Wermeille, J. W. Kim, J. C. Woicik, C. S. Hellberg, H. Li, Appl. Phys. Lett. **90**, 221908 (2007).

[9] J. C. Woicik, E. L. Shirley, C. S. Hellberg, K. E. Andersen, S. Sambasivan, D. A. Fischer, B. D. Chapman, E. A. Stern, P. Ryan, D. L. Ederer, H. Li, Phys. Rev. B **75**, 140103 (2007).

[10] M. P. Warusawithana, C. Cen, C. R. Sleasman, J. C. Woicik, Y. L. Li, L. F. Kourkoutis, J. A. Klug, H. Li, P. Ryan, L. P. Wang, M. Bedzyk, D. A. Muller, L. Q. Chen, J. Levy, D. G. Schlom, Science **324**, 367 (2009).

[11] A. A. Demkov and A. B. Posadas, *Integration of Functional Oxides with Semiconductors* (Springer, 2014).

[12] J. H. Ngai, D. P. Kumah, C. H. Ahn, F. J. Walker, Appl. Phys. Lett. **104**, 062905 (2014).

[13] M. Jahangir-Moghadam, K. Ahmadi-Majlan, X. Shen, T. Droubay, M. Bowden, M. Chrysler, D. Su, S. A. Chambers, J. H. Ngai, Adv. Mater. Int. **2**, 1400497 (2015).

[14] Z.-H. Lim, K. Ahmadi-Majlan, E. D. Grimley, Y. Du, M. Bowden, R. Moghadam, J. M. Lebeau, S. A. Chambers , J. H. Ngai, J. Appl. Phys. (2017).

[15] J. H. Ngai, K. Ahmadi-Majlan, J. Moghadam, M. Chrysler, D. Kumah, F. J. Walker, C. H. Ahn, T. Droubay, Y. Du, S. A. Chambers, M. Bowden, X. Shen, D. Su, J. Mater. Res. **32**, 249 (2017).

[16] R. M. Moghadam, Z. Y. Xiao, K. Ahmadi-Majlan, E. D. Grimley, M. Bowden, P. V. Ong, S. A. Chambers, J. M. Lebeau, X. Hong, P. V. Sushko, J. H. Ngai, Nano Lett. **17**, 6248 (2017).

[17] M. D. McDaniel, T. Q. Ngo, A. Posadas, C. Q. Hu, S. R. Lu, D. J. Smith, E. T. Yu, A. A. Demkov, J. G. Ekerdt, Adv. Mater. Int. **1**, 1400081 (2014).

[18] S. Y. Chen and L. W. Wang, Chem. Mater. **24**, 3659 (2012).

[19] S. A. Chambers, Y. Du, R. B. Comes, S. R. Spurgeon, P. V. Sushko, Appl. Phys. Lett. **110**, 082104 (2017).

[20] Certain commercial equipment, or materials are identified in this document. Such identification does not imply recommendation or endorsement by the National Institute of Standards and Technology, nor does it imply that the products identified are necessarily the best available for the purpose.

[21] L. Jones, H. Yang, T. J. Pennycook, M. S. J. Marshall, S. Van Aert, N. D. Browning, M. R. Castell, P. D. Nellist, Advanced Structural and Chemical Imaging **1**, 8 (2015).

[22] O. L. Krivanek, M. F. Chisholm, V. Nicolosi, T. J. Pennycook, G. J. Corbin, N. Dellby, M. F. Murfitt, C. S. Own, Z. S. Szilagyi, M. P. Oxley, S. T. Pantelides, S. J. Pennycook, Nature **464**, 571 (2010).

[23] C. Ohphus, Adv. Struct. Chem. Imaging **3**, 13 (2017).

[24] G. Kresse and J. Furthmuller, Phys. Rev. B **54**, 11169 (1996).

[25] G. Kresse and J. Hafner, Phys. Rev. B **49**, 14251 (1994).

[26] P. E. Blochl, Phys. Rev. B **50**, 17953 (1994).

[27] J. P. Perdew, A. Ruzsinszky, G. I. Csonka, O. A. Vydrov, G. E. Scuseria, L. A. Constantin, X. L. Zhou, K. Burke, Phys. Rev. Lett. **100**, 136406 (2008).

[28] S. A. Chambers, Y. Du, Z. Zhu, J. Wang, M. J. Wahila, L. F. J. Piper, A. Prakash, J. Yue, B. Jalan, S. R. Spurgeon, D. M. Kepaptsoglou, Q. M. Ramasse, P. V. Sushko, Phys. Rev. B **97**, 245204 (2018).

[29] X. H. Sang, Y. Xie, M. W. Lin, M. Alhabeb, K. L. Van Aken, Y. Gogotsi, P. R. C. Kent, K. Xiao, R. R. Unocic, Acs Nano **10**, 9193 (2016).

[30] G. Stone, C. Ophus, T. Birol, J. Ciston, C. H. Lee, K. Wang, C. J. Fennie, D. G. Schlom, N. Alem, V. Gopalan, Nat. Comm. **7**, 12572 (2016).

[31] S. B. Mi, C. L. Jia, V. Vaithyanathan, L. Houben, J. Schubert, D. G. Schlom, K. Urban, Appl. Phys. Lett. **93**, 101913 (2008).

[32] D. J. Smith, H. W. Wu, S. R. Lu, T. Aoki, P. Ponath, K. Fredrickson, M. D. McDaniel, E. Lin, A. B. Posadas, A. A. Demkov, J. Ekerdt, M. R. McCartney, J. Mater. Res. **32**, 912 (2017).

[33] S. Tanuma, C. J. Powell, D. R. Penn, Surf. Interface Anal. **43**, 689 (2011).

[34] C. J. Powell and S. Tanuma, in *Hard X-ray Photoelectron Spectroscopy (HAXPES)*, edited by J. C. Woicik (Springer, 2016), p. 111.

[35] A. Jablonski, J. Phys. D **48**, 075301 (2015).

[36] A. Jablonski, Surf Sci. **667**, 121 (2018).

[37] S. A. Chambers, T. C. Droubay, C. Capan, G. Y. Sun, Surf Sci. **606**, 554 (2011).

[38] T. Kendelewicz, S. Kaya, J. T. Newberg, H. Bluhm, N. Mulakaluri, W. Moritz, M. Scheffler, A. Nilsson, R. Pentcheva, G. E. Brown, J. Phys. Chem. C **117**, 2719 (2013).

[39] K. A. Stoerzinger, W. T. Hong, E. J. Crumlin, H. Bluhm, M. D. Biegalski, Y. Shao-Horn, J. Phys. Chem. C **118**, 19733 (2014).

[40] E. Landemark, C. J. Karlsson, L. S. O. Johansson, R. I. G. Uhrberg, Phys. Rev. B **49**, 16523 (1994).

[41] M. Wojtaszek, J. Lis, R. Zuzak, B. Such, M. Szymonski, Appl. Phys. Lett. **105**, 042111 (2014).

[42] M. Wojtaszek, R. Zuzak, S. Godlewski, M. Kolmer, J. Lis, B. Such, M. Szymonski, J. Appl. Phys. **118**, 185703 (2015).

[43] A. M. Kolpak and S. Ismail-Beigi, Phys. Rev. B **85**, 195318 (2012).

[44] S. A. Chambers, unpublished.

**Table 1 – Core-level binding energies and valence and conduction band offsets (in eV)**

|        | Ge 3d$_{5/2}$* | Sr 3d$_{5/2}$ | Ti 2p$_{3/2}$ | $\Delta E_V$ (a) | $\Delta E_V$ (b) | $\overline{\Delta E_V}$ | $\Delta E_C$ (c) |
|--------|----------------|---------------|---------------|------------------|------------------|-------------------------|------------------|
| 3 u.c. | 29.65(5)       | 133.85(2)     | 459.03(2)     | 3.05(10)         | 2.85(10)         | 2.95(10)                | 0.36(10)         |
| 6 u.c. | 29.65(5)       | 133.80(2)     | 459.06(2)     | 3.00(10)         | 2.88(10)         | 2.94(10)                | 0.35(10)         |
| 9 u.c. | 29.70(5)       | 133.82(2)     | 459.11(2)     | 2.97(10)         | 2.88(10)         | 2.92(10)                | 0.33(10)         |
| 32 u.c.| 29.93(5)       | 133.66(2)     | 459.03(2)     | 2.76(10)         | 2.69(10)         | 2.72(10)                | 0.13(10)         |

Notes:
* Values determined for the interface Ge layer taken from the band bending analysis described in the text.
(a) $\Delta E_V = \left(E_{Sr3d5/2} - E_{Ge3d5/2}\right)_{int} + \left(E_{Ge3d5/2} - E_V\right)_{Ge} - \left(E_{Sr3d5/2} - E_V\right)_{STO}$
(b) $\Delta E_V = \left(E_{Ti2p3/2} - E_{Ge3d5/2}\right)_{int} + \left(E_{Ge3d5/2} - E_V\right)_{Ge} - \left(E_{Ti2p3/2} - E_V\right)_{STO}$
(c) $\Delta E_C = \Delta E_V + E_g^{Ge} - E_g^{STO}$

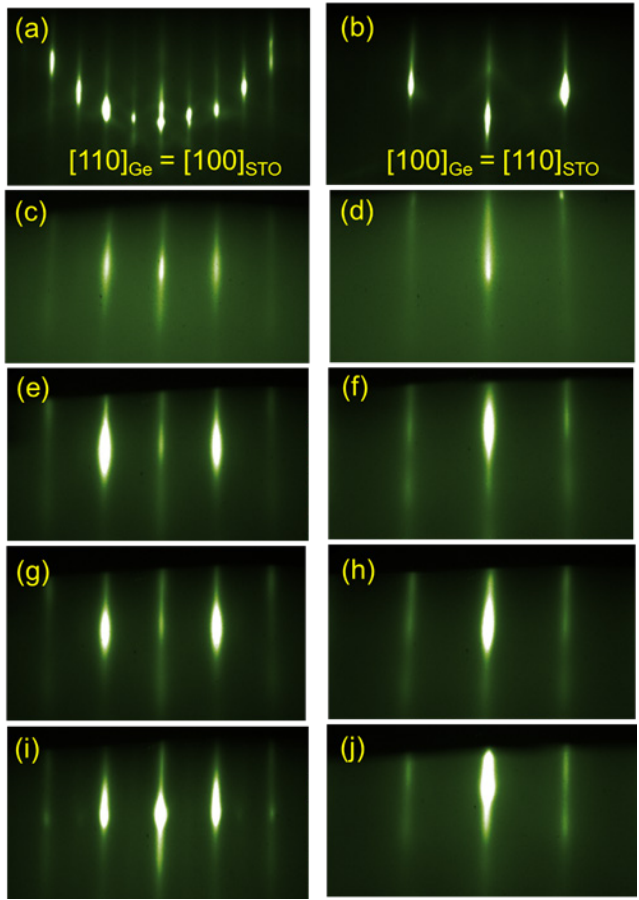

FIG. 1. (Color online) RHEED patterns in high-symmetry zone axes for clean *p*-Ge(001)-(2x1) (a & b), and STO/*p*-Ge(001) heterojunctions with STO thicknesses of 3 (c & d), 6 (e & f), 9 (g & h) and 32 (i & j) u.c.

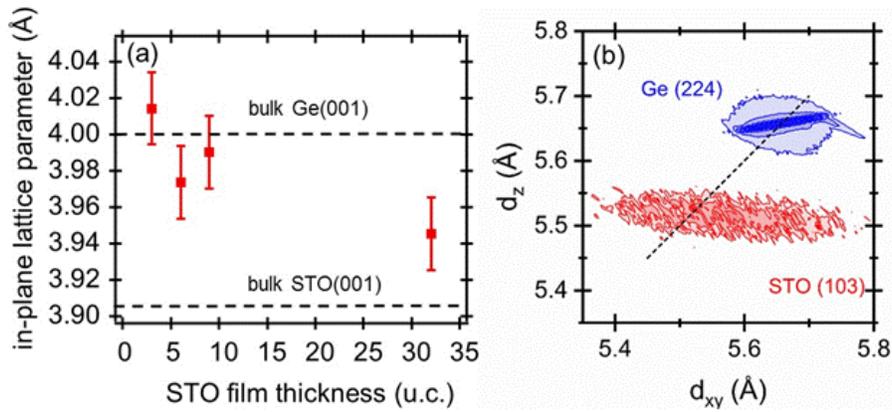

FIG. 2. (Color online) (a) In-plane lattice parameters extracted from streak spacings in RHEED patterns measured in both zone axes shown in Fig. 1: (b) XRD direct space map for 32 u.c. STO/Ge(001).

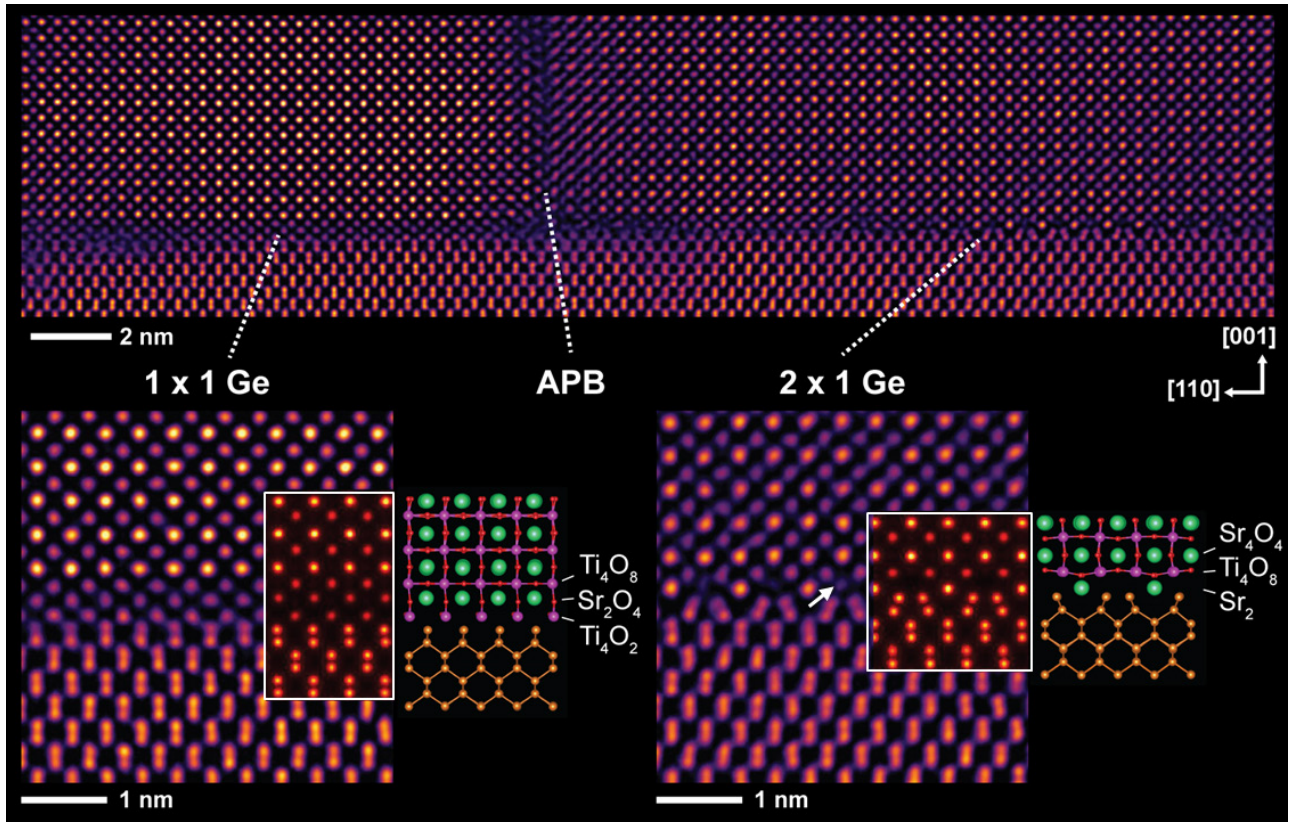

FIG. 3. (Color online) Cross-sectional STEM-HAADF image of the 32 u.c. film along the Ge [110] zone axis showing two dominant interface configurations—termed "1×1 Ge" and "2×1 Ge"—overlaid with multi-slice simulations (inside white rectangles) and relaxed first-principles models (to the right of the simulations). This image is the result of a non-rigid alignment of 10 frames followed by lattice filtering, as described in the methods. The compositions for the different layers are those of the super cells used in the first-principles calculations. Atoms: O = red, Ti = magenta, Sr = green, and Ge = orange.

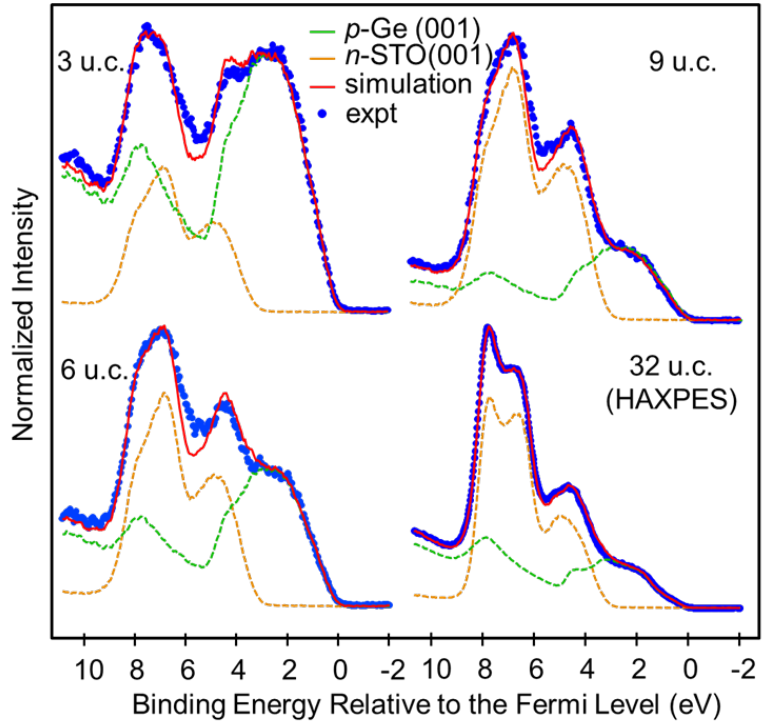

FIG. 4. (Color online) VB XPS measured at normal emission with $h\nu = 1487$ eV for 3, 6, and 9 u.c. STO/$p$-Ge(001), and HAXPES with $h\nu = 5930$ eV for 32 u.c. STO/$p$-Ge(001), along with fits to linear combinations of spectra for bulk $p$-Ge(001) and $n$-STO(001) in which the scaling factors and the VBO are free parameters.

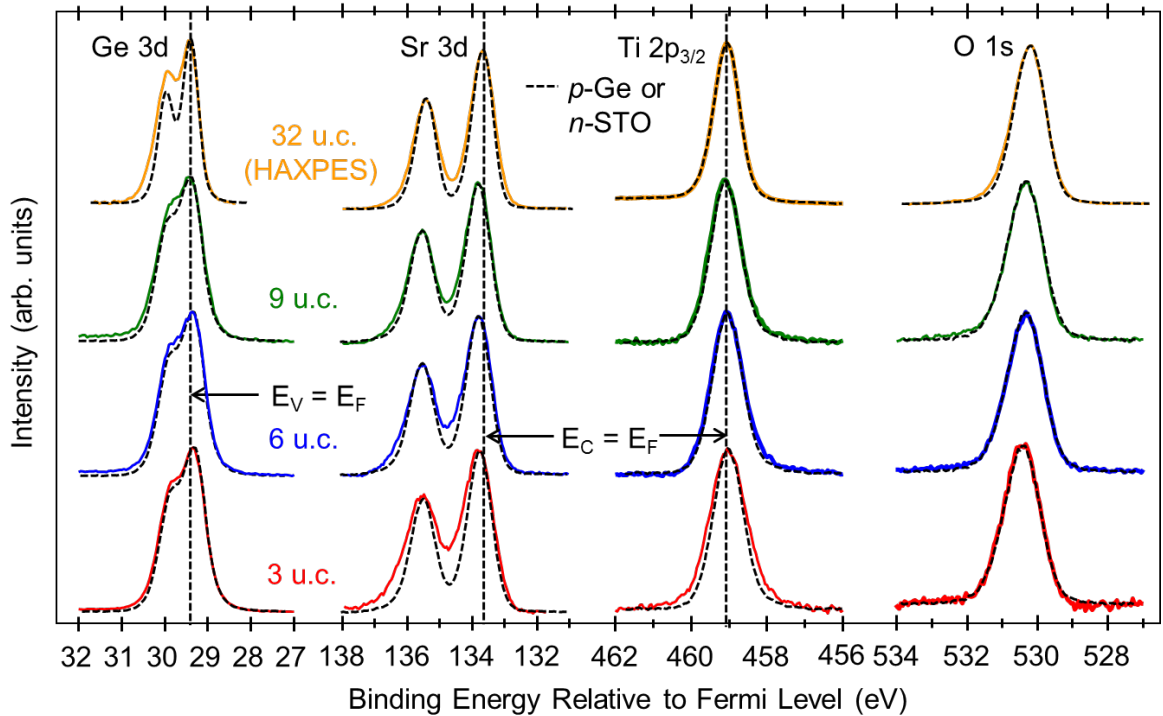

FIG. 5 (Color online) Core-level spectra for STO/$p$-Ge(001) heterojunctions with STO film thicknesses of 3, 6, 9 and 32 u.c. overlaid with reference (dashed) spectra measured for pure, flat-band $p$-Ge(001) and STO(001). Spectra were measured with an x-ray energy of either 1487 eV (3, 6 and 9 u.c.) or 5930 eV (32 u.c.). The dashed vertical lines indicate the binding energies that would reveal the Ge VBM being degenerate with the Fermi level ($E_V = E_F$), and the STO CBM being degenerate with the Fermi level ($E_C = E_F$).

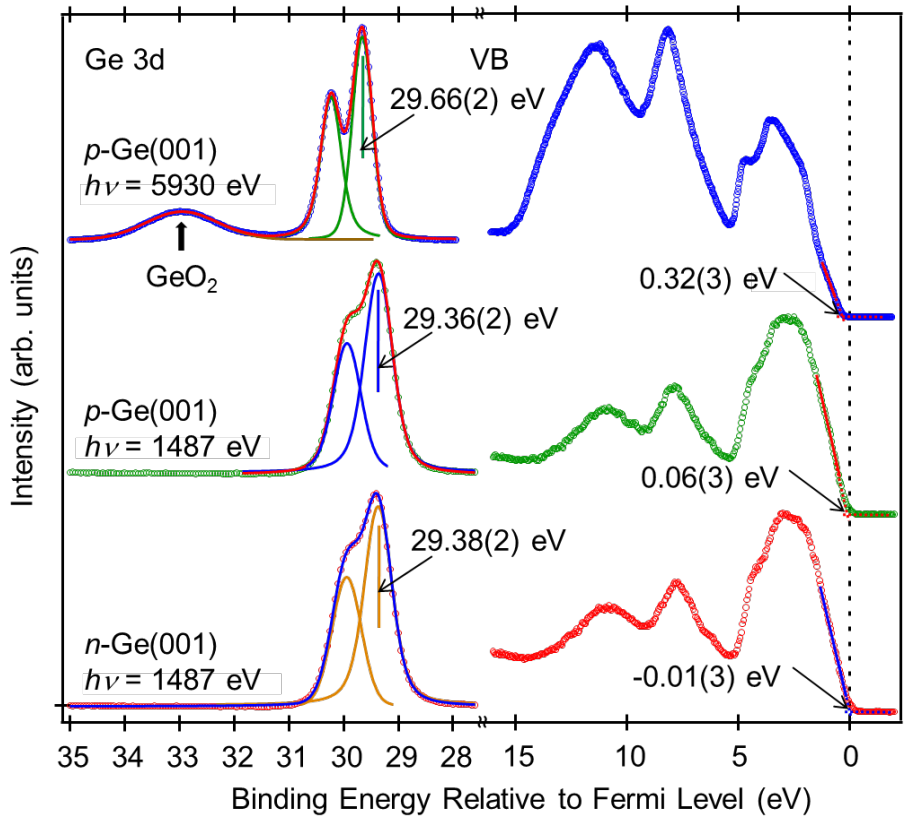

FIG. 6 (Color online) Ge 3d and VB spectra for clean *n*- and *p*-Ge(001)-(2x1) measured with AlKα x-rays, and for as-received *p*-Ge(001) with a thin native oxide measured with hard (~6 keV) x-rays.

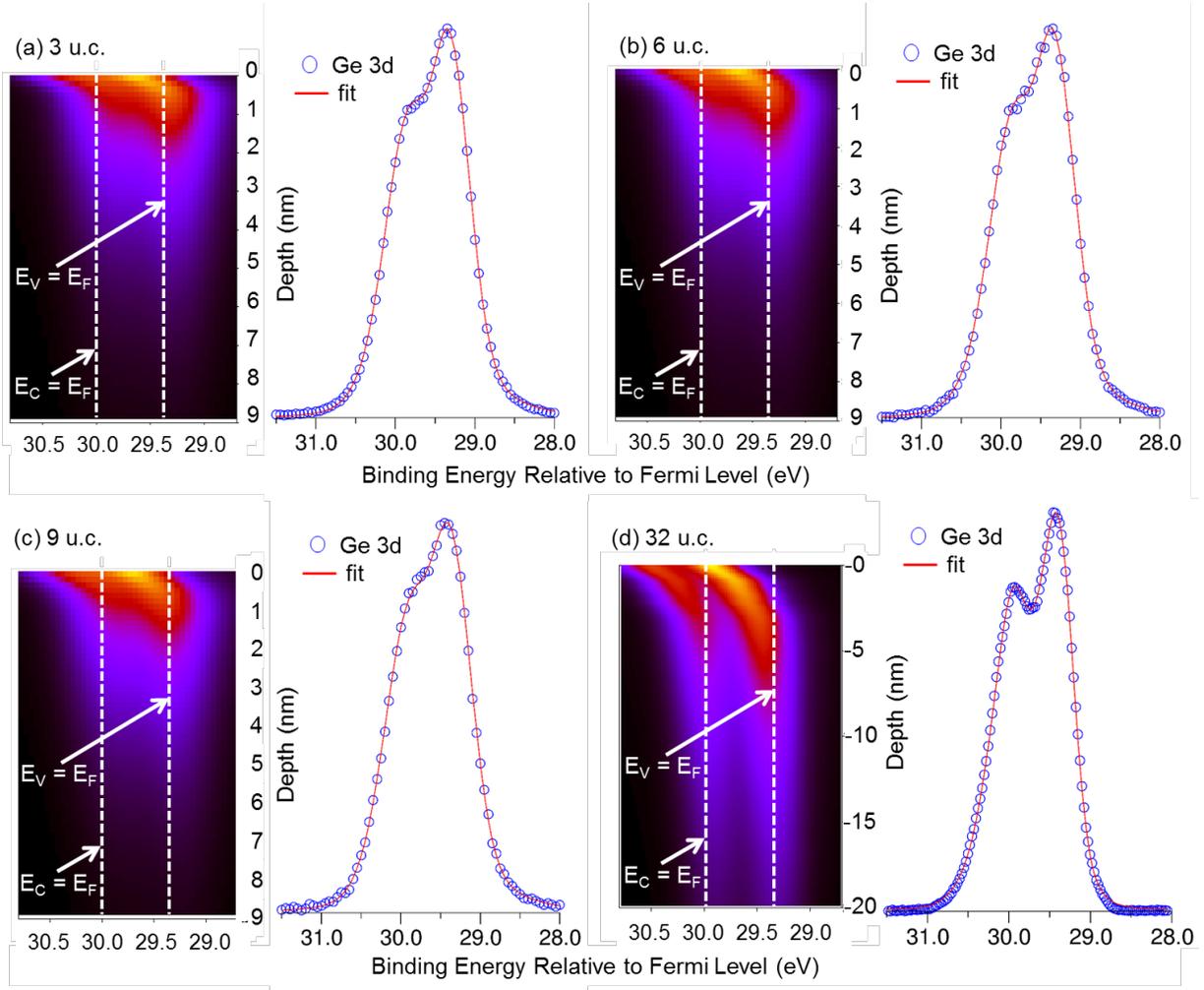

FIG. 7. (Color online) Fits of AlKα excited Ge 3d line shapes for 3, 6, 9 and 32 u.c. STO/$p$-Ge(001) heterojunctions to linear combinations of a model flat-band Ge 3d spectrum assigned to each layer within the probe depth. The weighting factors for the different layers are $\exp(-z/\lambda)$ where the electron attenuation length ($\lambda$) is 2.1 nm in (a-c) and 8.6 nm in (d) and $z$ is the depth below the interface. The total number of Ge planes included in the simulations are 100 for 3, 6 and 9 u.c., and 400 for 32 u.c, sufficient to include ~99.9% of the total Ge 3d intensity. The number of Ge planes subjected to binding energy variation ($m$) are 60 for 3, 6 and 9 u.c., and 120 for 32 u.c. The smoothing factor ($p$) used for all fits is 0.05. The best-fit spectrum resulting from summing over layers is overlaid with the experimental heterojunction spectrum on the right side of each panel, and the individual spectra are shown as contour maps on the left. The dashed vertical lines indicate the binding energies that would reveal the Ge CBM and VBM being degenerate with the Fermi level ($E_C = E_F$ and $E_V = E_F$).

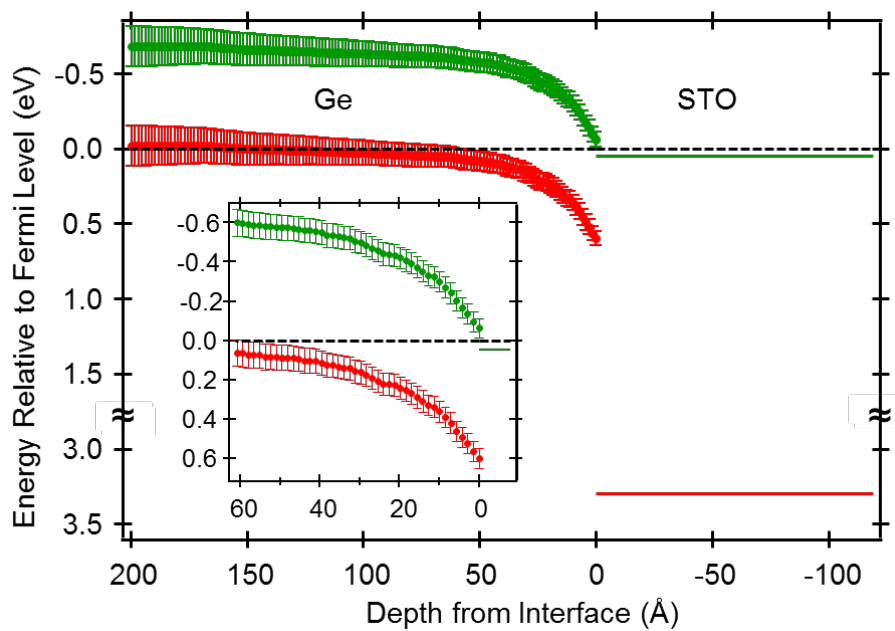

FIG. 8. Energy diagram for 32 u.c. STO/*p*-Ge(001) based on fitting the Ge 3d, Sr 3d and Ti 2p spectra in Figs. 5 & 7, and the resulting binding energies shown in Table I. The red circles (Ge) and red line (STO) are extracted from the photoemission data. The green circles (Ge) and green line (STO) are obtained from the red data by adding the band gaps of Ge and STO. The inset shows an expanded view of the Ge band edge behavior in the first 60 Å below the interface.

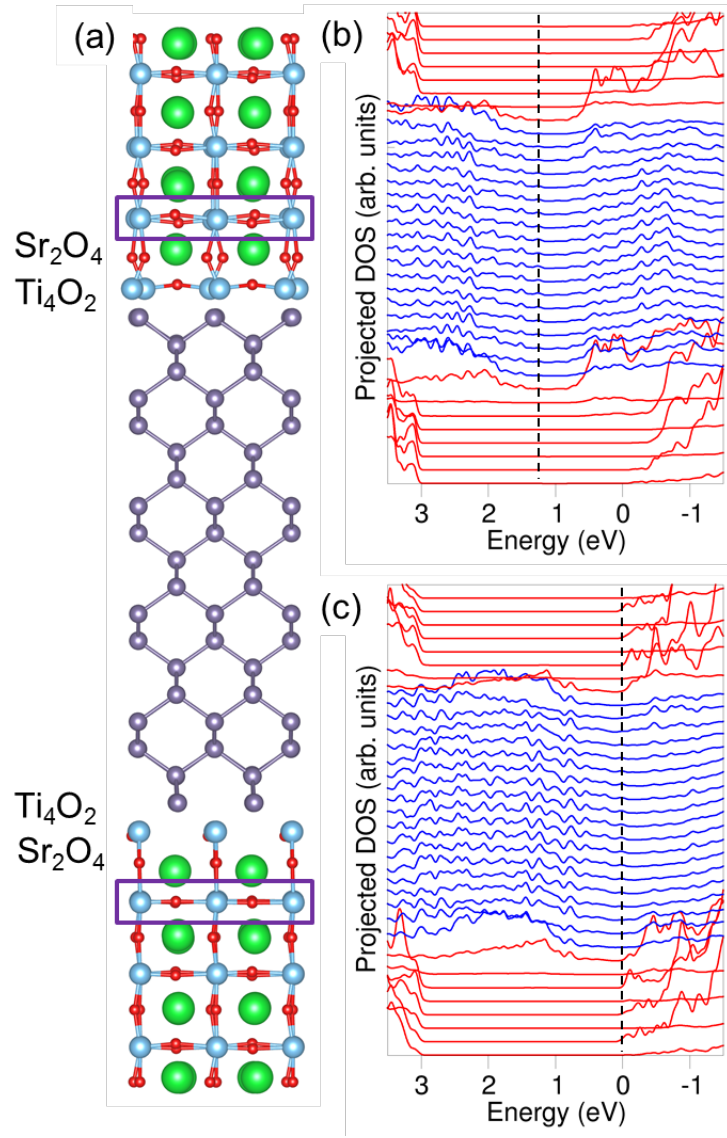

FIG. 9. (Color online) Layer-projected DOS for the "1×1 Ge" structure shown in Fig. 3 with (c) and without (b) O vacancies in the TiO$_2$ layer marked with a rectangle in (a). The dashed vertical lines mark the positions of the Ge VBM.

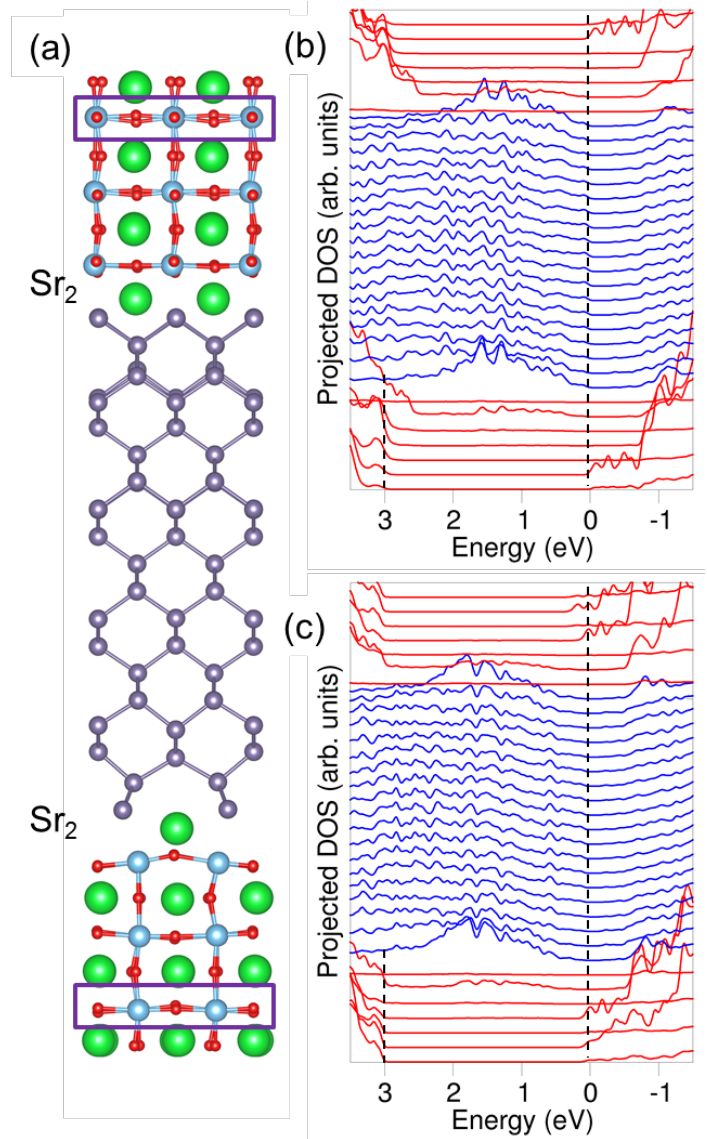

FIG. 10. (Color online) Layer-projected DOS for the "2×1 Ge" structure shown in Fig. 3 with (c) and without (b) O vacancies in the TiO$_2$ layer marked with a rectangle in (a). The dashed vertical lines mark the positions of the Ge VBM.